\documentclass[showpacs,twocolumn,aps,preprintnumbers,letterpaper]{revtex4}
\usepackage{amsmath,amssymb}
\usepackage{epsfig}
\usepackage{graphicx}
\usepackage{amsmath}
\usepackage{slashed}
\usepackage{amsfonts}
\usepackage{epstopdf}
\usepackage{color}

\newcommand{\be}{\begin{equation}}
\newcommand{\ee}{\end{equation}}
\newcommand{\bear}{\begin{eqnarray}}
\newcommand{\eear}{\end{eqnarray}}
\newcommand{\ba}{\begin{array}}
\newcommand{\ea}{\end{array}}

\def\be{\begin{eqnarray}}
\def\ee{\end{eqnarray}}
\def\bea{\be}
\def\eea{\ee}

\def\roughly#1{\mathrel{\raise.3ex\hbox{$#1$\kern-.75em%
\lower1ex\hbox{$\sim$}}}}

\begin{document}

\title{Entanglement in Reggeized Scattering using AdS/CFT}
\author{Yizhuang Liu}
\email{yizhuang.liu@sju.edu.cn}
\affiliation{Tsung-Dao Lee Institute, Shanghai Jiao University, Shanghai, 200240, China}
\author{Ismail Zahed}
\email{ismail.zahed@stonybrook.edu}
\affiliation{Department of Physics and Astronomy, Stony Brook University, Stony Brook, New York 11794-3800, USA}



\date{\today}
\begin{abstract}
The eikonalized parton-parton scattering amplitude at large $\sqrt{s}$ and large impact parameter,
is dominated by the exchange of a hyperbolic surface in  walled AdS. Its analytical continuation yields
a worldsheet instanton that is at the origin of the Reggeization of the amplitude and a thermal-like
quantum entropy ${\cal S}_T$. We explicitly construct the entangled density matrix following from 
the exchanged surface, and show that its von-Neumann entanglement entropy ${\cal S}_E$ 
coincides with the thermal-like entropy, i.e. ${\cal S}_T={\cal S}_E$.  The ratio of the
entanglement entropy to the transverse growth of the exchanged surface is similar to the Bekenstein 
entropy ratio  for a black-hole, with a natural definition of saturation and the on-set of chaos
in high energy collisions.  The largest  eigenvalues of the entangled density matrix obey a 
cascade equation in rapidity, reminiscent of non-linear QCD evolution of wee-dipoles  at low-x
and weak coupling.  We suggest that the largest eigenvalues  describe
the probability distributions of wee-quanta at low-x and strong coupling 
that maybe measurable at present and future  pp  and ep colliders.
\end{abstract}


\maketitle

\setcounter{footnote}{0}


\section{Introduction}

Entanglement in quantum mechanics is still one of the most subtle concept that permeates our description
of the quantum world. The canonical example is the entangled Einstein-Podolsky-Rosen  pair whereby
the measurement of one of the state in the pair forces the state of the partner. This conundrum has recently 
been revisited in many areas of physics, ranging from low-dimensional quantum critical systems~\cite{CRI,CRIX} to 
wormholes in gravity~\cite{JUAN,KRIS}.  

Recently the holographic principle  was used to derive the entanglement of boundary conformal field 
theories in terms of pertinent  area of finite dimensional surfaces in bulk~\cite{Ryu:2006bv}, reviving the idea
that the entanglement entropy bears similarities with the Bekenstein entropy for black holes~\cite{Srednicki:1993im}. 
These relationships are important in our understanding of the concept of information storage or loss
whether in quantum mechanics or around a black hole.

Current high multiplicity pp collisions at collider energies display 
rapid collectivization~\cite{CMS}, an indication of early entropy deposition and thermalization. 
This leads us to ask about the origin of this fast scrambling of information in the prompt phase of the process. 
One of  the purpose of this letter is to show that  parton-parton scattering at large $\sqrt{s}$ is  highly entangled,
with an entanglement entropy matching  the thermodynamical  entropy initially discussed in~\cite{Stoffers:2012mn}.
Entanglement entropies in the context of perturbative QCD evolution were recently 
discussed in~\cite{MANY,DIMA}.


Below we  briefly review the Reggeization of the parton-parton scattering at large $\sqrt{s}$, 
through the exchange of a minimal surface using the AdS/CFT correspondence. 
The transverse fluctuations on the surface are shown to be 
entangled with an  entropy that equals that of critical conformal field theories in lower
dimension.  The largest  eigenvalues of the entangled density matrix 
describe the probability distributions of wee-quanta at low-x  and strong coupling.

\begin{figure}[!htb]
 \includegraphics[height=50mm]{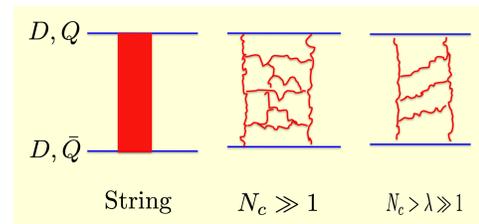}
 \caption{Schematic description of the string exchange in large $N_c$ (planar graphs) and holographic limit
 (ordered ladders) between a pair of dipoles ($DD$) or partons ($Q\bar Q$). See text.}
 \label{string}
\end{figure}

\section{Reggeized Scattering from AdS/CFT}

Elastic  hadron-hadron and lepton-hadron  collisions at large $\sqrt{s}$  are dominated by pomeron and reggeon exchanges
\cite{GRIBOV,Donnachie:1992ny}.  Perturbative QCD evolution describes these exchanges through
ordered gluon ladders~\cite{Kuraev:1977fs,Balitsky:1978ic}, while non-perturbative holographic descriptions suggest 
 string exchanges~\cite{SIN,JANIK,Basar:2012jb,MORE} or Reggeized bulk gravitons~\cite{POL}. Throughout, we will
 present the holographic string version. A brief review of this approach will be given in this section.

\subsection{Motivation}

At large $N_c$, QCD processes are dominated by planar graphs.  In the Pomeron limit
with $s\gg -t$, the scattering between a pair dipoles $DD$ or quarks $Q\bar Q$ is 
characterized by the exchange of planar graphs which can be regarded as the exchange of a closed string (Pomeron) or
an open string (Reggeon). In the holographic limit of a large number of colors and strong coupling $N_c\gg \lambda\gg  1$,
the exchanged gluons are qualitatively ordered as suggested in~\cite{SHURYAKZAHED}. 

\begin{figure}[!htb]
 \includegraphics[height=4cm]{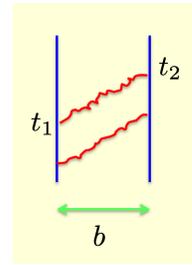}
 \caption{Ordered exchange at strong coupling.}
 \label{coulomb}
\end{figure}

For an intuitive understanding of this ordering, it is best to consider the original Maldacena$^\prime$s modified
Coulomb law~\cite{MALDACENA}. For that, consider the ordered gluon contribution for the potential between two heavy quarks as illustrated
in Fig.~\ref{coulomb}. In Feynman gauge where the retardation is manifest, the ordered gluons contribute to the potential as

\be
\label{VB}
V(b)= -\frac{\lambda}{4\pi^2}  \int \frac {dt_{12}}{t_{12}^2+b^2}
\ee
In the abelian case, the interaction takes place at all virtualities 
with typically the dominant and large times $\Delta t_{12}\sim b$, leading to the standard Coulomb 
potential $V(b)\sim -\lambda/b$. At strong coupling, the non-abelian modified Coulomb law 
is seen to be dominated by  short time exchanges $\Delta t_{12}\sim b/\sqrt{\lambda}\ll b$, with 
(\ref{VB}) giving

\be
\label{VBB}
V(b)\sim -\frac{\lambda \, \Delta t_{12}}{b^2}\sim -\frac{\sqrt{\lambda}}b
\ee
At strong coupling,
the coherence captured by the potential can only build if the exchanged non-abelian gluons travel super-luminally,  for otherwise they will
undergo multiple splitting and lose coherence because of the large gauge coupling $\lambda$. These rapid  exchanges  are suggestive of the ordering in
Figs.~\ref{string},\ref{coulomb}. We emphasize the qualitative and intuitive character of this argument, which does not allow for fixing the
overall coefficient in (\ref{VBB}) for instance.

\subsection{Eikonal amplitude}

In the eikonal limit, the probe and target partons support Wilson lines
running along the light cone and sourcing gluon fields~\cite{Nachtmann:1991ua}. Specifically, 
the parton-parton scattering amplitude at large $N_c$, is given by
 ($t=-{\bf q}^2$)

\begin{eqnarray}
\label{XDD2}
{\cal T}_{ij, kl}(s,t) 
= -2is\int d {\bf b} \ e^{i {\bf q}\cdot {\bf b}} \, {\bf WW}_{ij,kl} \ , \label{dipdip6}
\end{eqnarray}
with the connected Wilson loop correlator

\be
\label{XDD3}
{\bf WW}_{ij,kl}\equiv \langle {\bf W}_{ij}(C_1) {\bf W}_{kl}(C_2) \rangle_{\rm con}
\ee
traced over colors, and subject to the normalization  $\langle {\bf W}_{ii}\rangle=N_c$.
The Wilson lines ${\bf W}_{ij}$ are evaluated along  $C_{1,2}$  on the light cone at fixed separation 
$b=|{\bf b}|$ 

\be
{\bf W}_{ij}(C_{1,2})=\left({\bf P}\, e^{ig\int_{C_{1,2}}A}\right)_{ij}
\ee
as illustrated in Fig.~\ref{kinematics} following~\cite{Nachtmann:1991ua,SIN,JANIK}. 
The averaging in (\ref{XDD3}) is over the Yang-Mills gauge fields.  The integrand in (\ref{XDD2}) is the
impact parameter representation of the scattering amplitude in the s-channel.

Vacuum gauge invariance allows the decomposition of (\ref{XDD2}) into a singlet and octet contribution

\be
\label{DXX4}
{\cal T}_{ij, kl}={\cal T}_0\delta_{ij}\delta_{kl}+{\cal T}_{N_c^2-1}\,T^{a}_{ij}T^a_{kl}
\ee
where $T^a$ are the generators of SU($N_c$) in the fundamental representation.  Each of the amplitude
in (\ref{DXX4}) can be obtained by a pertinent closing of the $C_{1,2}$ contours at infinity,  leading to

\be
&&{\cal T}_0=\frac 1{N_c^2}\left<{\bf WW}_{ii,jj}\right>-1\nonumber\\
&&{\cal T}_0+\frac {N_c^2-1}{2N_c}{\cal T}_{N_c^2-1}=\frac 1{N_c}\left<{\bf WW}_{ij,ji}\right>-1
\label{T0T}
\ee
Both the singlet and octet amplitudes are gauge invariant and can be assessed using perturbative or non-perturbative
arguments. We choose to evaluate them using non-perturbative arguments in the context of
holography which we now present.

\subsection{Holography}

In the holographic limit,
these gauge invariant amplitudes are dominated by string exchanges. 
In leading order, the correlator of Wilson lines involve surface exchange 
in a slice of AdS with a metric

\be
ds^2=\frac {R^2}{z^2}(-dt^2+dz^2+dx^idx^i)
\label{5}
\ee
for $0\leq z\leq z_0$ with $D_\perp=3$. The invariant correlators  in (\ref{T0T})
are dominated by the minimal surface  attached to $C_{1,2}$~\cite{MALDACENA}

\be
\label{3}
{\bf WW}\sim  e^{-\sigma_T{\cal A}_{\rm min}}\equiv e^{-S_{\rm min}}
\ee
The singlet amplitude  involves the exchange of a closed surface with the topology of a cylinder, 
while the octet amplitude  involves the exchange of an open surface with the topology
of a disc, as  discussed in~\cite{SIN,JANIK,Basar:2012jb,MORE}. 
Here, we present a simplified analysis where the inelasticity carried by the exchanged surfaces is encoded
in a generic world-sheet instanton irrespective of the topology of the surface.

\begin{figure}[!htb]
 \includegraphics[height=50mm]{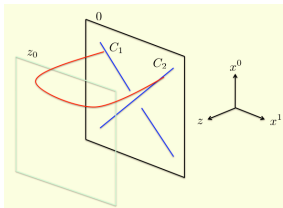}
 \caption{Wilson lines with the attached string  $0\leq z\leq z_0$.}
 \label{kinematics}
\end{figure}

Consider the open string exchange. For  large impact parameter $b$, the 
extremal surface is composed  of two straight strips joined by a surface at 
$z=z_0$ as shown in Fig.~\ref{kinematics}. The two straight strips contribute about 1 in (\ref{T0T})
with  the  normalization 
$\langle {\bf W}_{ii}\rangle=N_c$. 
To assess the joining surface at $z=z_0$ where the metric is nearly flat, we use the Polyakov action in the {\bf conformal} gauge
with mostly positive  Minkowski signature ($a=\tau, \sigma$)

\be
S=\frac {\sigma_T}2 \int_0^Td\tau \int_0^1d\sigma\,\partial_a x\cdot \partial_ax +{\rm b.c.}
\label{3x}
\ee
with the string tension 

\be
\sigma_T=\frac{R^2}{(2\pi \alpha_{\mathbb X}^\prime z_0^2)}\sim \frac{\sqrt{\lambda}}{(2\pi z_0^2)}
\ee

Our treatment of the closed (${\mathbb X}=\mathbb P$) and open (${\mathbb X}=\mathbb R$)
topologies will be similar except for: 1/ an adjustment of the  string tension by choosing
 
 \be
 \label{LL}
 \alpha^\prime_{\mathbb P}=\frac 12 \alpha^\prime_{\mathbb R}\equiv \alpha^\prime\,,
 \ee
 and  2/ an additional $\frac 1s$ suppression of the scattering amplitude (\ref{XDD2}) for he open surface 
 exchange due to the running quark lines on the open boundary~\cite{YEE}. Throughout we will set $R=z_0$.

For large $b$, the  boundary conditions at
$z=0$ transfer almost unchanged to $z=z_0$. These boundaries are straight
lines  with rapidity angles  $\chi/2$ and $-\chi/2$ for $\sigma=0,1$ respectively, with $\chi={\rm ln}s\gg 1$. 
At $\sigma=0$ we have~\cite{Basar:2012jb}

\be
&&\cosh(\chi/2)\partial_\sigma x^0 +\sinh(\chi/2)\partial_\sigma x^1=0\,,\nonumber\\
&&\sinh(\chi/2)\partial_\tau x^0+\cosh(\chi/2)\partial_\tau x^1=0\,,
\label{bdr1}
\ee
 and similarly at $\sigma=1$.  The extremal solution to (\ref{3x}) with
 $\partial_a^2x=0$ and subject to (\ref{bdr1})
 at $z=z_0$, is the hyperbolic surface ($\bar\sigma=\sigma-\frac 12$)

\be
&&(x^0, x^1, x^\perp, z)=\nonumber\\
&&\bigg(\frac b\chi \cosh\left(\chi\bar\sigma\right)\sinh(\chi\tau), \frac b\chi\sinh\left(\chi\bar\sigma\right)\sinh(\chi\tau), b\sigma, z_0\bigg)\nonumber\\
\label{SOL1}
\ee
The  induced world-sheet metric associated to (\ref{SOL1}) is conformal, 

\be
ds_W^2=b^2\cosh^2(\chi \tau)(-d\tau^2+d\sigma^2)
\ee
which is consistent (a posteriori)   with the gauge choice in (\ref{3x}).
It is free of the wormhole discussed in~\cite{KRIS}.
Using the analytical continuation $\tau\rightarrow i\tau$,
we have 

\be
ds_W^2\rightarrow b^2\cos^2(\chi \tau)(d\tau^2+d\sigma^2)
\ee
 which describes the conformal world-sheet 
of an  $^{\prime\prime}$instanton$^{\prime\prime}$ with period $T_P=2\pi/\chi\ll 1$  and finite action

\be
S_{\rm min}=\sigma_T\,\int_0^{T_P}b^2{\rm cos}^2(\chi\tau)\,d\tau\int_0^1\,d\sigma=\frac 12 \sigma_T (bT_P)b\nonumber\\
\label{action}
\ee

\begin{figure}[!htb]
 \includegraphics[height=6cm]{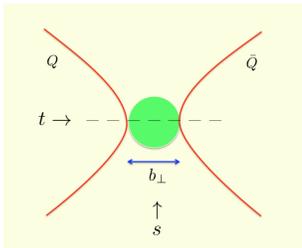}
  \caption{Eikonal  scattering with the instanton exchange.}
  \label{fig_long}
\end{figure}

From (\ref{LL}), it follows that 
the world-sheet  instanton contribution for the Pomeron is twice that of
the Reggeon with $\sigma_{\mathbb P}=2\sigma_{\mathbb R}$ and $S_{\mathbb P}=2S_{\mathbb R}$.
The closed surface exchange can be thought as two glued open surface exchanges.
 In Fig.~\ref{fig_long} we give an illustration of the geometrical relationship between the 
worldsheet instanton and the hyperbolic surface sustained by the nearly eikonal trajectories.
A more thorough characterization of this  instanton and
its relation to the Schwinger mechanism on the world-sheet can be found in~\cite{Basar:2012jb}
(see section IIID).

\subsection{Reggeized amplitude}

Following the AdS/CFT correspondence, we insert 
(\ref{action}) into (\ref{3x}) and define $\beta=bT_P$,   to obtain  
(\ref{3}) as

\be
{\bf WW}&\sim& e^{-S_{\rm min}-S_{\rm 1loop}}\nonumber\\
&=&e^{-\frac 12\sigma_T\beta b+\frac{D_\perp}{12}\,\chi}= e^{-\frac{b^2}{2\alpha_{\mathbb X}^\prime \chi}+\frac {D_\perp}{12}\chi}
\label{SWW}
\ee
We have included the 1-loop  quantum correction restricted to the world-sheet instanton 
strip $b\times \beta$. For $\chi\gg 1$, the strip is highly elongated $b\gg \beta$ and {\bf periodic} in $\beta$.
The 1-loop contribution is dominated by the Casimir energy or Luscher term~\cite{LUSCHER}

\be
S_{\rm 1loop}&=&\frac{D_\perp}2\,{\rm ln}\,{\rm det}( -\partial^2_\perp)\nonumber\\
&=&-\frac{\pi D_\perp}{6}\frac b\beta=
-\frac{D_{\perp}}{12}\chi
\label{1loop}
\ee


Inserting (\ref{SWW})  in (\ref{XDD2}) and carrying the 
transverse Fourier transform yields  the Reggeized scattering amplitude
${\cal T}_{\mathbb X}\sim is^{\alpha_{\mathbb X} (t)}$
($R=z_0$)

\be
\alpha_{\mathbb P}(t)=&&1+\frac{D_\perp}{12}+\frac {\alpha^\prime}2 t\nonumber\\
\alpha_{\mathbb R}(t)=&&1-1+\frac{D_\perp}{12}+{\alpha^\prime} t
\ee
in agreement with~\cite{Basar:2012jb} (see section IV) for the Pomeron, and with~\cite{YEE} for the Reggeon.
Note the $-1$ from the additional $\frac 1s$ suppression in this channel as stated earlier.

\subsection{Warped Gribov diffusion}

To exponential accuracy, (\ref{SWW}) happens to be exactly the tachyon-mode contribution to the  closed string propagator subject
to the  twisted boundary
conditions (\ref{bdr1}). Specifically, in the Pomeron channel the exact tachyon contribution is~\cite{MORE,SZ}

\be
\label{KTB}
\mathbb K_0(t_\chi,b)=\left(\frac 1{4\pi t_\chi}\right)^{\frac {D_\perp}2}\,e^{-\frac{b^2}{4t_\chi}+\frac {D_\perp t_\chi}{6\alpha^\prime}}\sim {\bf WW}
\ee
with the rapidity playing  the role of time 

\be
\label{TIME}
t_\chi=\frac{\alpha^\prime}2\chi \equiv D_{\mathbb P}\chi
\ee
and  $D_{\mathbb P}$ playing the role of a diffusion constant.
 For large rapidity $\chi\gg 1$, all other string excitations  are suppressed. 
The world-sheet instanton in (\ref{SWW}) captures semi-classically the tachyon contribution in (\ref{KTB}).

(\ref{KTB})   embodies
the famed Gribov diffusion, 

\be
\label{DIFFUSION}
\partial_{t_\chi}\mathbb K_0 +(M_0^2-\nabla_\perp^2)\mathbb K_0=0
\ee
with the tachyon mass $M_0^2=-D_\perp/6\alpha^\prime$.  It acts as a {\bf source} term in 
 the diffusion process. The Pomeron intercept, the Luscher term  and the tachyon mass
are intimatly related in our analysis.  
It is now clear, that the effects of warping
 amount to a warped Gribov diffusion

\be
\label{DIFFUSIONW}
\partial_{t_\chi}\mathbb K_0 +
\left(M_0^2-\frac 1{\sqrt{g_\perp}}\partial_\mu g_\perp^{\mu\nu}\sqrt{g_\perp}\partial_\nu\right)\mathbb K_0=0
\ee
with $g_\perp$ the   transverse AdS-metric.
The transverse  directions include the holographic z-direction, so that $\mathbb K_0(t,b)\rightarrow \mathbb K_0(t, b, z)$
as $z=z_0$ is now relaxed. For AdS the solution to (\ref{DIFFUSIONW}) can  be obtained in closed form.
Specifically, for $z\ll b$ and ${\bf D_\perp=3}$ we have~\cite{MORE} (see Eq. 38)

\be
\label{BFKLX}
\frac 1{zz_0} {\mathbb K_0(t_\chi,b,z)} \approx 
\frac{e^{(\alpha_{\mathbb P}(0)-1))}}{(4\pi t_\chi)^{\frac 32}}\frac{2z}{z_0b^2}
{\rm ln}\left(\frac{b^2}{zz_0}\right)\,e^{-\frac 1{4t_\chi}{\rm ln^2}\left(\frac{b^2}{zz_0}\right)}\nonumber\\
\ee
Modulo the string parameters $\alpha_{\mathbb P}$ and $D_{\mathbb P}$,
(\ref{BFKLX}) is identical  to Mueller$^\prime$s  BFKL density of wee-dipoles of size $z$
in onium-onium scattering in the 1-Pomeron approximation at weak coupling~\cite{MUELLER}
(see Eq. 8 in section II).   

Remarkably, (\ref{DIFFUSIONW}) interpolates between the scattering amplitude of the soft Pomeron 
(\ref{KTB}) and the hard Pomeron (\ref{BFKLX}) in impact parameter space for exactly ${\bf D_\perp=3}$. Therefore, it is natural
to interpret the string zero point fluctuations in the exchanged instanton world-sheet as wee-dipoles
at strong coupling, much like Mueller$^\prime$s wee-dipoles at weak coupling. We will return to this point below.


\section{Thermal-like Entropy}

The  exchanged instanton  period or 
tunneling time,  
plays the role of an inverse temperature for the zero point fluctuations on the induced
world-sheet

\be
\beta=bT_P=\frac{2\pi b}\chi
\ee
This temperature is kinematical in origin, as it  arises from the rapidity $\chi$ 
of the colliding pairs for fixed impact parameter $b$.
The larger the rapidity and/or smaller the impact
paramer, the shorter the tunneling time or higher the temperature.

This physical observation is important. It shows that the  Casimir energy or Luscher term in (\ref{1loop}) is
 the  {\bf free energy} $ {\cal F}_T$ of $D_\perp$  massless bosons confined to a  
1-dimensional box of length   $b$ at temperature $1/\beta$, 

\be
\beta {\cal F}_T& =&
D_{\perp}\int \frac{bdp}{2\pi}\ln (1-e^{-\beta |p|})\nonumber\\&=&-\frac{\pi D_{\perp}}{6}\frac b\beta\nonumber\\
&=&S_{\rm 1loop}
\label{1loopX}
\ee
The zero point fluctuations on the instanton world-sheet are {\bf thermal-like}
Hence,  the exchanged instanton plus zero-point motion carry  a thermal entropy ${\cal S}_T$ 
 that follows from standard thermodynamics

\be
{\cal S}_T&=& \frac{\beta\partial  {\cal F}_T}{\partial{\rm ln}\beta}\nonumber\\
&=&D_\perp\int \frac{bdp}{2\pi}\frac {2\beta|p|}{e^{\beta|p|}-1}\nonumber\\
&=&\frac{D_{\perp}}{6}\chi=2(\alpha_{\mathbb P}(0)-1)\chi
\label{THERMO}
\ee
in agreement with the initial observation in~\cite{Stoffers:2012mn,SZ}.
 The thermal-like entropy (\ref{THERMO}) per unit rapidity $\chi$
 is fixed by the Pomeron intercept $\alpha_{\mathbb P}(0)$. It is  at the origin of the rise
 of the scattering cross section  and ultimatly the multiplicities in high energy scattering
 as discussed in~\cite{Stoffers:2012mn,SZ}.

\section{Entanglement Entropy}

In this section we show that  the string thermal-like
 entropy ${\cal S}_T$ in (\ref{THERMO})  is {\bf identical} to the entanglement
 von Neumann  entropy ${\cal S}_E$ following from the blocked density matrix of the transverse part of the
 exchanged string.
 The longitudinal part of the string  freezes out due to
Lorentz contraction at large rapidity. Throughout this section we will set $\alpha^\prime =\frac 14$, and reinstate it
when needed by inspection.

\subsection{Transverse Hamiltonian}

The transverse fluctuations  at the origin of (\ref{1loop}-\ref{THERMO}) are associated to 
the Polyakov action (\ref{3x}) with $x\rightarrow x_\perp$ and  fixed end-points  around the 
hyperbolic configuration (\ref{SOL1}),

\be
S_\perp=\frac {\sigma_T}2 \int_0^Td\tau \int_0^1d\sigma\ \left({\dot{x}_\perp}^2-{x_\perp^\prime}^2\right)+{\rm b.c.}
\label{3xx}
\ee
The action density   can be thought as that of a collection of $N$  strings
connected by identical springs for $z=z_0$, and discretized as follows
~\cite{Karliner:1988hd,Bergman:1997ki,EARLY}

\be 
\frac{1}{N}    \sum_{k=0}^{N}  \left( \dot{x}_\perp^i (k) \right)^2 - \frac{1}{N}     \sum_{k=1}^N \left(  \frac{x_\perp^i(k) - x_\perp^i (k-1)}{\frac{\pi}{N}}    \right)^2 
\label{DIS}
\ee
where the summation over $i=1, ..., D_\perp$ is subsumed.
 Note that (\ref{DIS}) reduces to (\ref{3xx}) as $N\rightarrow\infty$. 
(\ref{DIS}) describes $N$ coupled harmonic oscillators in $D_\perp$ dimensions,  with a transverse Hamiltonian

 \be
 \frac 2N \mathbb H_\perp=\frac 12 \sum_{k=0}^{N} \left( p^i_k \right)^2 +\frac 12 \sum^N_{k,l=1}
 x^i_k\mathbb K_{kl}x^i_l
 \label{HTT}
 \ee
where $\mathbb K$ is a  banded matrix 
 
 \be
 \mathbb K_{kl}=\frac 4{\pi^2}\left(2\delta_{kl}-\delta_{k,l+1}-\delta_{k,l-1}\right)
 \label{BAND}
 \ee
 with positive eigenvalues.  
 Ignoring warping at large $b$, the ground state wave function of (\ref{HTT}) is
 
 \be
 \Psi[x]=\left(\frac{|\Omega |}{\pi^N}\right)^{\frac{D_\perp}4}e^{-\frac 12 \sum_{k,l=1}^Nx_k^i\Omega_{kl}x_l^i}
 \label{GROUND}
 \ee
 where $\Omega$ is the square root of $\mathbb K$.

 Since $\mathbb K$ is real symmetric,
 it diagonalizes by ortogonal rotation with $\mathbb K=U^{\dagger} K_DU$ and $\Omega=U^{\dagger}\sqrt{K_D}U$. 
The eigenvalues and eigen-vectors of $\mathbb K$ are respectively

\be
\lambda_k=&&\frac{8}{\pi^2}\bigg(1-\cos \frac{\pi k}{2p+1}\bigg)\nonumber\\
\alpha_k^{n}=&&\sqrt{\frac{2}{2p+1}}\sin \frac{\pi k n}{2p+1}
\ee
with $k$ labeling the eigenvalues and $n$ labeling the entries,  $k,n=1,2,..2p$. The matrices $U,\Omega$ 
can be found in explicit form, with $U_{kn}=\alpha_k^n$ and

 \be
 \Omega_{mn}=\sum_{s=\pm} \frac{s\,C}{\cos \frac{\pi (m-sn)}{2p+1}-\cos \frac{\pi}{4p+2}}
 \label{MATRIX}
 \ee 
with $C$ an overall unimportant constant. 
Given $\Omega_{mn}$, the derivation of  the entanglement entropy is essentially an exercise in 
the diagonalization of nested Gaussians as in~\cite{Srednicki:1993im}.

  \subsection{Density matrix}

  The transverse string density matrix is $\Psi[x]\Psi^*[x^\prime]$.
 To quantify the entanglement of the string bits in transverse
 space, we  follow Srednicki~\cite{Srednicki:1993im} and define the entanglement density

\be
\rho_E[\bar x, {\bar x}^\prime]=\int d[\underline{x}]\,\Psi[\underline{x},\bar x]\Psi^*[\underline{x};{\bar x}^\prime]
\label{DENS}
\ee
where we used the notation $[x]\rightarrow [\underline x;\bar x]$ with ${\rm dim}\,\underline x=n$ and ${\rm dim}\,\bar x=N-n$.
The positive eigenvalues of (\ref{DENS}) follow by diagonalization

\be
\int d[{\bar x}^\prime] \,\rho_E[\bar x, {\bar x}^\prime]\,\varphi_l[{\bar x}^\prime]=p_l\varphi_l[\bar x]
\label{DXX}
\ee
The entanglement entropy is the Von-Neumann entropy for the transverse string 

\be
{\cal S}_E(n,N)=-\sum_{l=0}^\infty p_l\,{\rm ln}(p_l)
\label{SnN}
\ee

\subsection{Von-Neumann entropy}

For the 2 limiting cases $n=1$ and $n=N/2$  (\ref{SnN}) can be obtained in closed form.
For general $n$ the eigenvalues $p_l$ can only be obtained numerically.
 For that,  we fix the end points through the boundary condition $x_{N+1}=x_1=0$.
Without loss of generality, we set $N=2p$ and subdivide  $N$ into 

\be
[N]=[n]\cup  [N-n]
\ee
The entanglement entropy between the subsystem with size $[n]$ and 
the one with size $[N-n]$ can be calculated by splitting the matrix $\Omega$
in (\ref{MATRIX}) as 

\be
B_{N,n}=&&\Omega_{m \bar m}, m \in (1,..n), \bar m \in (n+1,..N)
\ee
and defining the squared matrix $\tilde \beta$ through 

\be
\beta_{N,n}=\Omega_{N,n}^{-\frac{1}{2}}B_{N,n}\Omega^{-1}_{N,N-n}B^{T}_{N,n}\Omega_{N,n}^{-\frac{1}{2}}
\equiv \Omega_{N,n}^{-\frac{1}{2}}\tilde \beta_{N,n}\Omega_{N,n}^{\frac{1}{2}}\nonumber\\
\ee
The corresponding  eigenvalue spectrum  follows from

\be
\tilde\beta_{N,n}v_{N,n,i}=\chi_{N,n,i}v_{N,n,i}
\ee
with $0\leq i\leq n$.
For each transverse dimension $1,...,D_\perp$, the eigenvalues of the entangled density matrix (\ref{DXX})  
are~\cite{Srednicki:1993im}

\be
p_l[N,n,i]=\left(1-\xi_{N,n,i}\right)\,\xi^l_{N,n,i}
\label{pnN}
\ee
with

\be
\xi_{N,n,i}=\frac{\chi_{N,n,i}}{\chi_{N,n,i}+\sqrt{1-\chi^2_{N,n,i}}}
\ee 
The entanglement entropy (\ref{SnN}) is then 

\be
&&{\cal S}_E(n,N)=-D_\perp\sum_{i=1}^{n}\sum_{l=0}^\infty p_l[N,n,i]\,{\rm ln}\, p_l[N,n,i]\nonumber\\
&&=-D_\perp\sum_{i=1}^{n}\left(\ln(1-{\xi_{N,n,i}})+\frac{{\xi_{N,n,i}}}{1-{\xi_{N,n,i}}}\ln {\xi_{N,n,i}}\right)\nonumber\\
\label{S1S2}
\ee

\begin{figure}[!htb]
 \includegraphics[height=50mm]{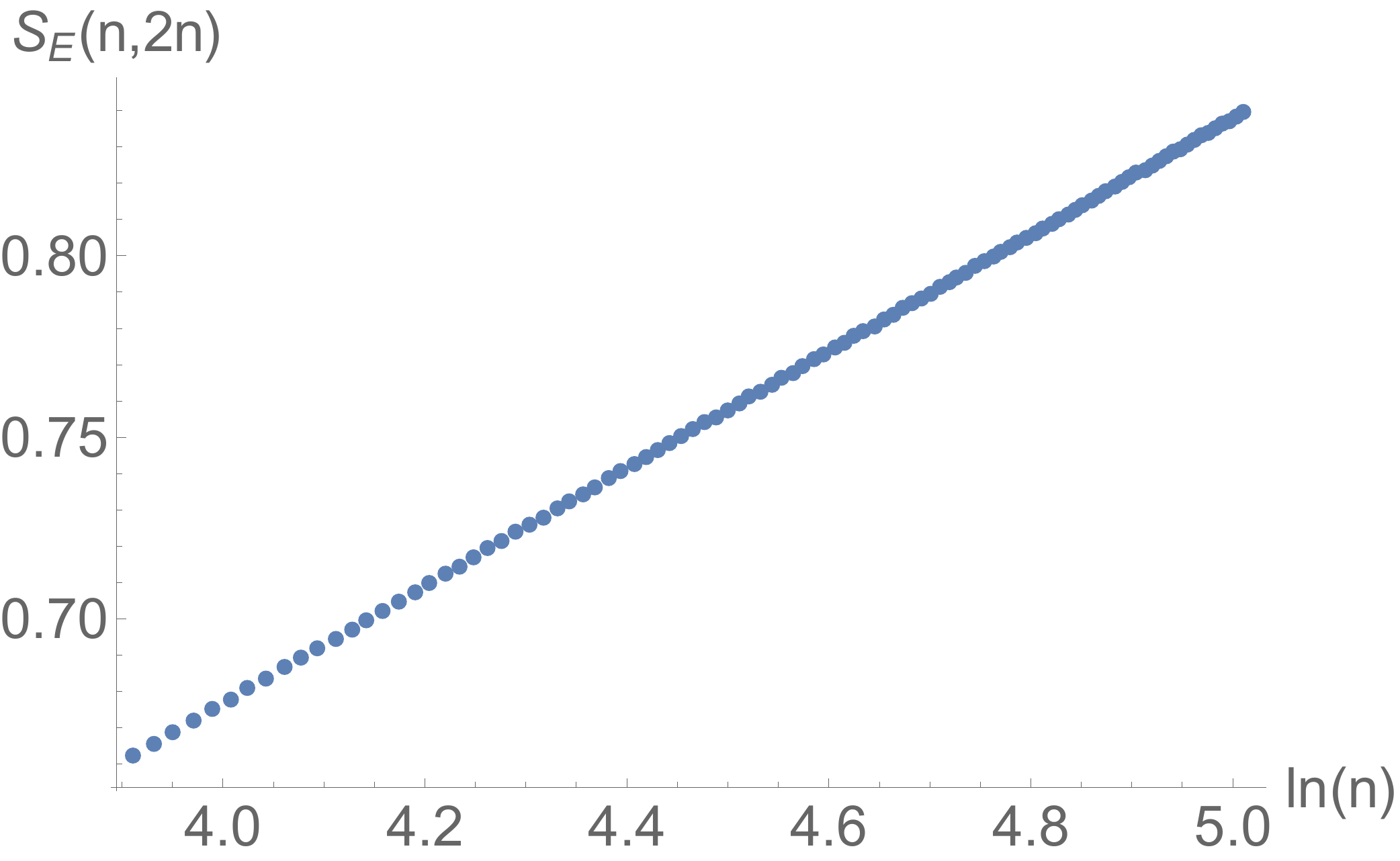}
  \caption{The entanglement entropy ${\cal S}_E(n, 2n)$ per $D_\perp$ versus ${\rm ln}(n)$ 
  in the range  $50\le n\le 150$ for a transverse string with fixed ends.}
  \label{entropy2}
\end{figure}

 In Fig.~\ref{entropy2} we show our results for 
${\cal S}_E(n, N=2n)$ versus ${\rm ln}(n)$ per $D_\perp$, in the range $ 50 \le n\le 250$,

 \be
 {\cal S}_E(n, N)=\frac {D_\perp}6\,{\rm ln}(n)\rightarrow
 \frac {D_\perp}6\,{\rm ln}\left(\frac{N}{\pi}\,{\rm sin}\bigg(\frac {n\pi}N\bigg)\right)
 \label{CFT}
 \ee
  Because of the mid-point symmetry of  the chain,  the last equation follows.
 We have checked that for the string with periodic boundary conditions, i.e. $x_{N+1}=x_1$,  (\ref{CFT}) is also recovered with $1/6\rightarrow 1/3$
(2 boundary points). (\ref{CFT})   is identical to the thermodynamical 
entropy (\ref{THERMO}) for $1\ll n\ll N$ with the identification of the rapidity $\chi={\rm ln}(n)$ (see below).
It is consistent with results from
conformal field theories and spin chains  with central charge $D_\perp$~\cite{CRI} (1 boundary point).

\subsection{Black-hole and chaos}

With increasing rapidity $\chi$, the exchanged string is longitudinally Lorentz contracted and transversely
more elongated and excited causing it to spread. The string transverse squared  size  is given by the averaging

\be 
R_\perp^2  (N)=  \frac{1}{N} \sum_{i=1}^{D_\perp}\sum_{k= 1}^{N-1}  \left< \left(x_k^i  \right)^2 \right> 
\label{RT0}
\ee 
with the probability distribution fixed by (\ref{GROUND})

\be
|\Psi[x]|^2=\left|\prod_{i=1}^{D_\perp}\prod_{k=1}^N\left(\frac{\sqrt{\lambda_k}}\pi\right)^{\frac 14}e^{-\frac{\sqrt{\lambda_k}}2x_k^{i2}}\right|^2
\ee
Each of the discretized string bit coordinates $x_k^i$ is normally distributed with probability $|\Psi[x]|^2$.
This gives rise to a random walk of the string bits along the chain in the transverse direction with fixed end-points. The transverse squared
size (\ref{RT0}) is 

\be 
R_\perp^2  (N)
&=& \frac{D_\perp}{2N}\sum_{k=1}^{N-1} \frac{1}{\sqrt{\lambda_k}}\nonumber\\
&\approx&  \frac{D_\perp}{4} \ln\, N \rightarrow  D_\perp\alpha^\prime {\rm ln}\,N
\label{RT0X}
\ee 
after reinstating the units $\frac 14\rightarrow\alpha^\prime$.

This  result is consistent with 
the unwarped Gribov diffusion for the Pomeron, since (\ref{KTB}) implies a transverse normal diffusive spread

\be
\label{RT0XX}
\left<b^2\right>=2D_\perp t_\chi=D_\perp \alpha^\prime \chi
\ee
with the averaging carried using (\ref{KTB}). A comparison of (\ref{RT0X}) with
(\ref{RT0XX}) shows that the number of string bits $N$ and the rapidity of the colliding
pair $\chi$,  are tied

\be
N=e^{\chi}
\label{RAPID}
\ee
 Recall that the rapidity $\chi$  and the Lorentz factor 
 $\gamma =1/\sqrt{1-\beta^2}$ in the relativistic limit are tied
 by $\chi={\rm ln}s=2{\rm ln}\gamma$.
The more we boost the string, the larger the rapidity $\chi$, the more string bits $N$ in the
transverse plane, the longer the intrinsic length $L=N\sqrt{\alpha^\prime}$ of the string.
 
 From (\ref{CFT}) it follows that the entanglement entropy growth is proportional to the
 squared transverse size of the string at the same resolution $n$

\be
\label{BEKE}
{{\cal S}_E(n)}=\frac {\pi R_\perp^2(n)}{6\pi\alpha^\prime} \,
\ee
This is reminiscent of the Bekenstein entropy  $S_{BH}$ for  a {\bf black-hole }
in relation to its  area $A_{BH}$~\cite{BKK}

\be
\label{BHH}
S_{BH}=\frac{A_{BH}}{4\,l_P^2}=\frac{\pi R_{BH}^2}{l_P^2}
\ee
In (\ref{BEKE})  the string length plays the role of an effective  Planck length $l_P$.
 
 Black holes are maximal scramblers. The correspondence between (\ref{BEKE}) and
 (\ref{BHH}) implies that  the entanglement entropy  {\bf density} saturates at  very large rapidities, 
 with a saturation momentum

\be
\label{QSS}
Q_S^2=\frac{S_E(n)}{\pi R_\perp^2(n)}=\frac 1{6\pi\alpha^\prime}\equiv \frac 1{l_P^2} 
\ee
This corresponds to 
1 unit of entanglement entropy per effective Planck area. For a string length
$\sqrt{\alpha^\prime}=0.1$ fm, the saturation momentum is $Q_S\approx 0.5$ GeV.

\begin{figure}[!htb]
 \includegraphics[height=6cm]{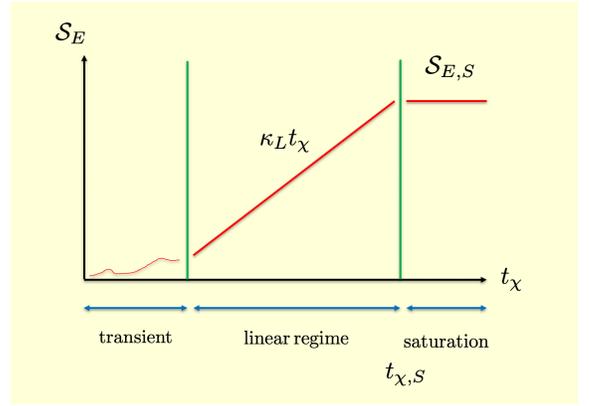}
  \caption{Typical production  of entanglement entropy by  evolution in rapidity $\chi$ 
  (also ${\rm ln}\frac 1x$)) or Gribov time
  $t_\chi$. See text.}
  \label{fig_set}
\end{figure}

We note that in terms of  the Gribov diffusion-like time $t_\chi$ in (\ref{TIME}), 
the entanglement entropy (\ref{THERMO}) grows linearly with $t_\chi$. 
This translates to a constant growth rate

\be
\label{RATE}
\frac{d{\cal S}_E}{dt_\chi}=\frac{D_\perp}{3\alpha^\prime}\equiv \kappa_L
\ee
In Fig.~\ref{fig_set} we show the typical evolution of the entanglement entropy with rapidity $\chi$, or low-x (see below) 
or Gribov time. 
After an initial transient, a linear regime takes place with a characteristic rate $\kappa_L$ ending in the saturation regime. 
The typical Gribov time for saturation $t_{\chi,S}\sim {\cal S}_{E,S}/\kappa_L$ is reached for a rapidity

\be
\label{SATXX}
\chi_S=\left(\frac {6}{D_\perp}\right)\,{\cal S}_{E,S}\rightarrow 
\frac 1{2(\alpha_{\mathbb P}(0)-1)}\,{\cal S}_{E,S}
\ee
For pp scattering, ${\cal S}_{E,S}$ can be estimated from (\ref{BHH}-\ref{QSS}) in the black disc limit 
with $\pi R_\perp^2= 1$ fm$^2$ and a  string length $\sqrt{\alpha^\prime}=0.1$ fm,  i.e.
${\cal S}_{E,S}=100/(6\pi)\approx 5$. For  $D_\perp=3$ the Pomeron intercept is  $\alpha_{\mathbb P}(0)=1.25$, this translates to 
$\chi_{S}=10.6$, hence a collision energy $\sqrt{s}/m_N\approx 200$. 
We note that our estimates
are sensitive to the numerical value of the intercept $\alpha_{\mathbb P}(0)$. 

In general chaotic systems, the growth rate of the entropy 
in physical time is usually bounded by the Kolmogorov-Sinai entropy rate $\lambda_L$, i.e. $|d{\cal S}/dt|\leq \lambda_L$, 
with $\lambda_L$ the sum of all positive  Lyapunov exponents~\cite{KOLMO}. It is a key measure of chaoticity.
The above analogy with the black hole suggests that the entanglement production rate (\ref{RATE}) is at the {\bf chaos bound}.


By contrast,  the classical entropy of the  string ${\cal S}_S$  grows faster. Specifically, for a string 
with N-string bits in $D_\perp$ dimensions, the total number of string states 
are  $N_S=D_\perp^N$, and its entropy is then

\be
{\cal S}_S={\rm ln}N_S=N{\rm ln}D_\perp=e^\chi\,{\rm ln}D_{\perp}
\ee
It is proportional to the total
mass or length of the string,  and grows faster than the quantum entanglement entropy $S_E$. 
In terms of the Gribov diffusion time, the corresponding rate is

\be
\label{RATEMORE}
\frac{d{\rm ln}{\cal S}_S}{dt_\chi}=\frac 1{D_{\mathbb P}}
\ee
with $D_{\mathbb P}$ the Pomeron diffusion constant. 
The  scrambling in Gribov time is  now seen to scale  logarithmically with ${\cal S}_S$.
A similar scaling of the scrambling in real time was noted for black holes~\cite{PRESKILL}.
The rate (\ref{RATEMORE}) violates the Kolmogorov-Sinai bound~\cite{KOLMO}. Such violations usually occur in a transient
regime,  as also noted for chaotic maps~\cite{KOLMO} (second reference). This rapid growth in the classical entropy of the string $S_S$ 
is expected to stop when string
self-interactions are included at saturation~\cite{EARLY}.

  \section{Wee-quanta}

We now  show that 
the largest eigenvalue of the entangled density matrix of the string carries most of the quantum collectivity,
and allows for a global characterization of the wee-dipoles or wee-quanta at strong coupling. This
discussion offers a complementary view of the Gribov diffusion discussed earlier, where the warped
amplitude for the worldsheet instanton plus zero point motion was shown 
to carry identical information to the distribution of  BFKL-like wee-dipoles, albeit at strong coupling.

\begin{figure}[!htb]
 \includegraphics[height=5cm]{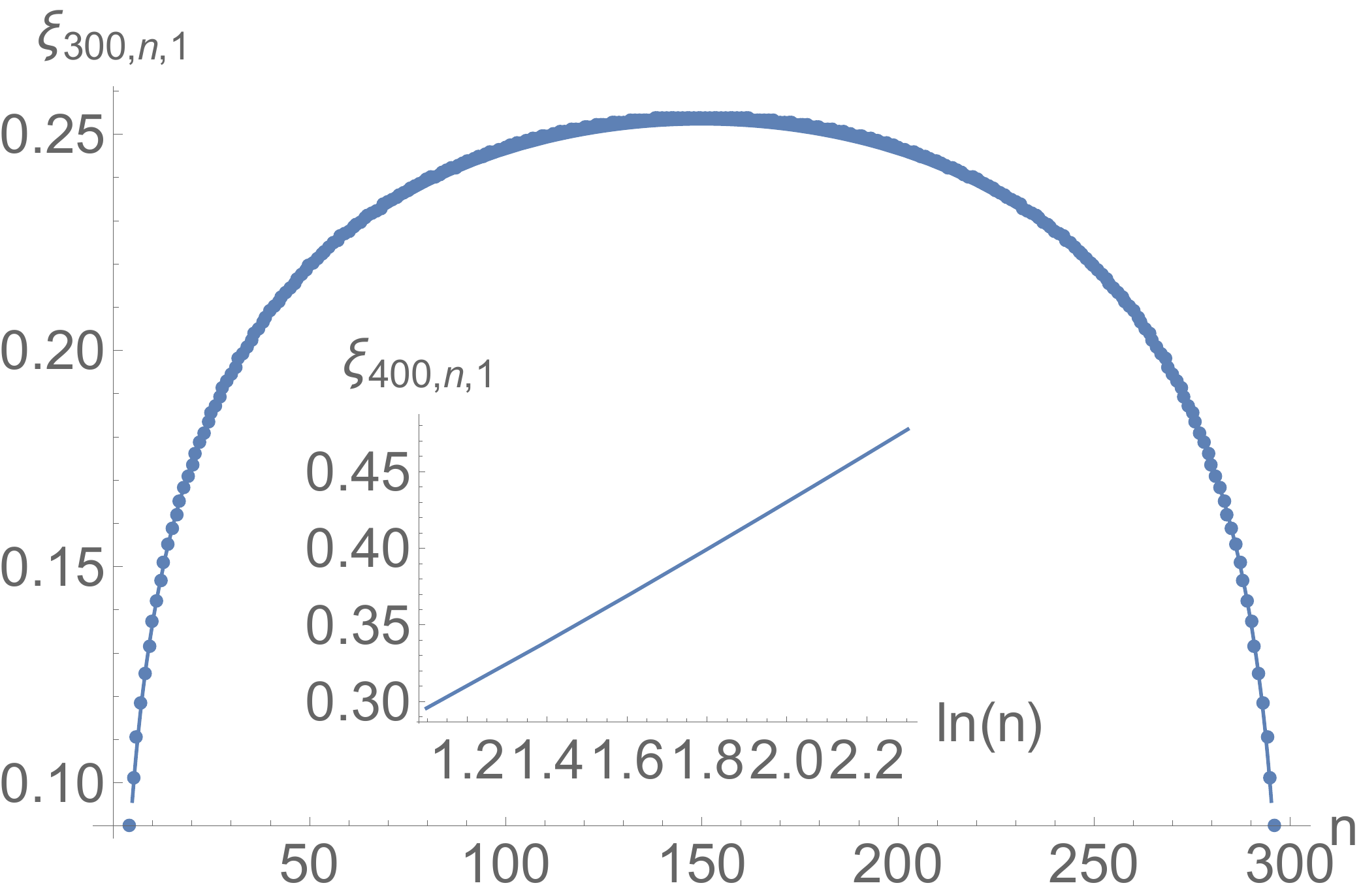}
  \caption{Largest eigenvalue $\xi_{300,n,1}$ versus $n$  in the full range $0< n<300$, with
  the mid-chain periodicity manifest: dots are the numerical results and the solid line is a best fit (\ref{LARGE}).
  The insert shows $\xi_{400,n,1}$  versus ${\rm ln}(n)$ in the range  $3\leq n\leq 10$.}
  \label{eigenvalue-n}
\end{figure}

\subsection{Largest eigenvalues}

\begin{figure}[!htb]
 \includegraphics[height=5cm]{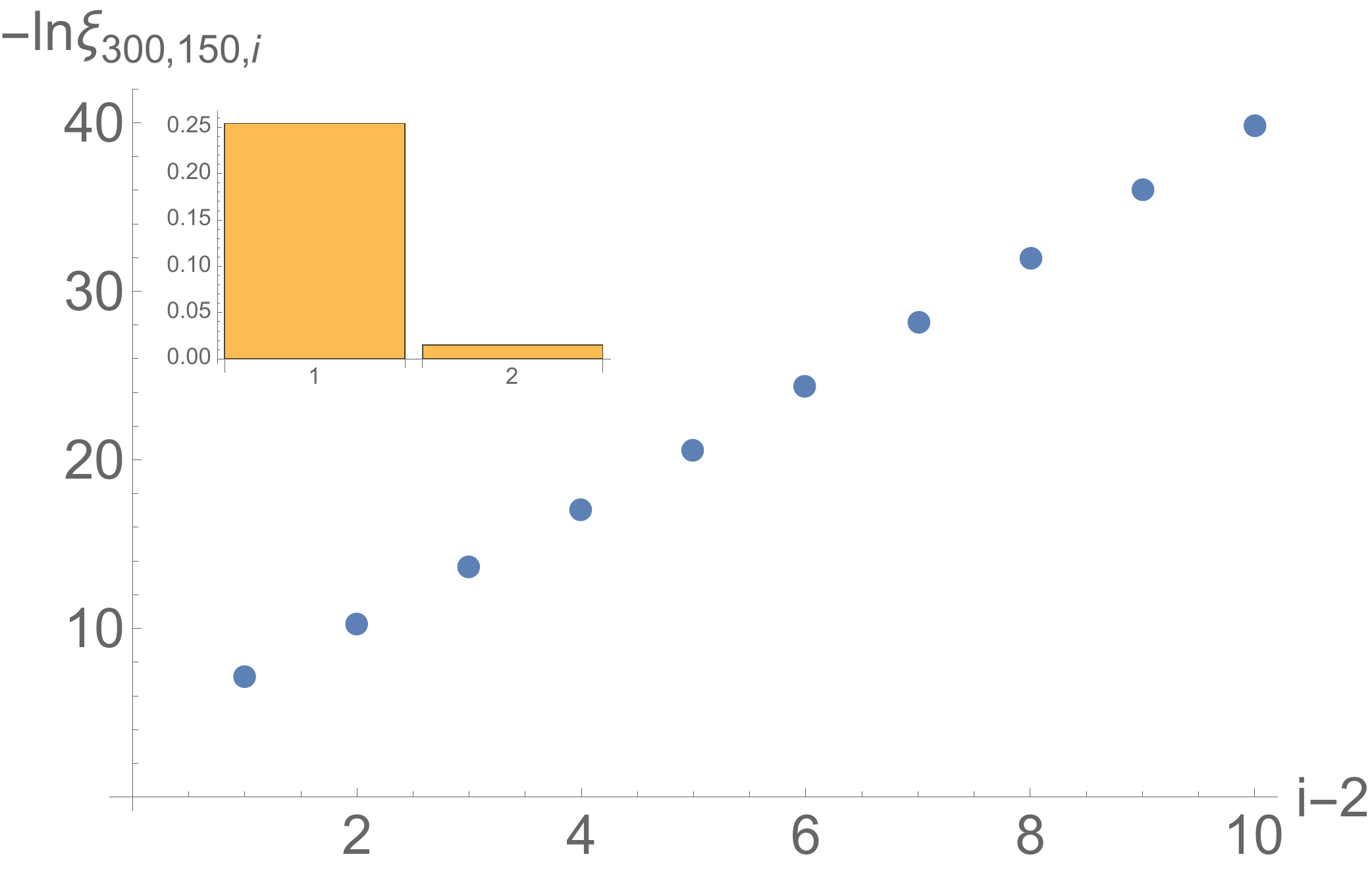}
  \caption{Distribution of the eigenvalues  $-\ln \xi_{300,150,i}$ versus $i-2$ for $3\le i\le 12$. The insert shows 
  the largest two eigenvalues $\xi_{300,150,i=1,2}$  on a linear scale.}
  \label{eigenvalue}
\end{figure}

The entanglement
entropy is dominated by the largest  two eigenvalues $\xi_{N,n,i=1,2}$ 
with  $\xi_{N,n,1}=0.155\,{\rm ln}(n)$ as shown in Fig.~\ref{eigenvalue-n},  which reproduces the entropy
 (\ref{CFT}) for  small ${\rm ln}(n)$. The eigenvalue distribution decreases  exponentially (Poisson),
 i.e. $ \xi_{2n,n,i>2}\approx e^{-a|i|}$,   as shown 
 in~Fig.~\ref{eigenvalue} for $2n=300$ and $a=3.65$.
The dependence of  $\xi_{300,n,1}$ on $n$ is shown by the dots on the semi-circle-like in
 Fig.~\ref{eigenvalue-n}, with the best fit  
 
 \be
 \xi_{N,n,1}=0.963\,
\bigg( 1-e^{-\Delta\,{\rm ln}
\left(\frac{N}{\pi}\,{\rm sin}\bigg(\frac {n\pi}N\bigg)\right)}\bigg)
 \label{LARGE}
 \ee
for $N=300$ and $\Delta=0.067$.
Using (\ref{LARGE}) in (\ref{pnN}) gives the dominant eigenvalues or probabilities
($n\ll N=300$)

\be
p_l[N,n,1]\approx e^{-\Delta\,{\rm ln}(n)}\,\left(1-e^{-\Delta\,{\rm ln}(n)}\right)^l
\label{pnN2}
\ee
with  $0.963\rightarrow 1$ for $D_\perp=1$. For large ${\rm ln}(n)$, the numerical analysis 
is more intensive, but  we expect $\Delta=0.067\rightarrow \frac 16$ as required by the entropy constraint (\ref{CFT}).

\subsection{Cascade equation}

In general, we have $D_{\perp}$ independent copies of string 
chains,  each with $l_{1,...,D_\perp}$, $n_{1,...,D_\perp}=n$. For fixed and common $n$ (rapidity), 
and fixed $l=l_1+...+l_{D_\perp}$, 
the largest eigenvalue  (\ref{pnN2}) is replaced by

\be
p_{l}[D_{\perp},N,n,1]=&&\sum_{l=l_1+..+l_{D_{\perp}}}\prod_{M=1}^{D_{\perp}}p_{l_M}[N,n,1]\nonumber\\=&&
\frac{(l+D_\perp-1)!}{l!D_\perp !}\nonumber\\
&&\times  e^{-D_\perp\Delta\,{\rm ln}(n)}\,\left(1-e^{-\Delta\,{\rm ln}(n)}\right)^l\nonumber\\
\label{GENERAL}
\ee 
which satisfies  a cascade equation in rapidity

\be
\frac {dp_l}{d{\rm ln}n}=-\Delta\,(l+1)\, p_l+\Delta\,(l+D_\perp -1)\,p_{l-1}
\ee
with $n=e^\chi$ following from (\ref{RAPID}).
In terms of the mean $\left<l\right>$, (\ref{GENERAL}) is a negative binomial  distribution

\be
\label{BINOM}
p_{l}[D_{\perp},N,n,1]=P^{NBD}(D_\perp, \left<l\right>-D_\perp, l)
\ee
For $D_\perp=1$ and modulo $\Delta$,  
(\ref{BINOM}) is identical  to  the probability to find $l$-wee-dipoles  inside a hadron 
at rapidity $\chi={\rm ln}(n)$  following from a model of non-linear QCD evolution
at weak coupling~\cite{DIMA}.

\subsection{Structure function at low-x}

Deep inelastic ep scattering at low-x is similar to pp scattering in the Pomeron regime.
The virtual photon exchange at large $Q^2$ in ep scattering, acts as a dipole of size
$1/Q$ scattering off the proton as a quark-diquark dipole, hence the similarity. 
 In the holographic limit, both involve the exchange of  a closed surface. To map the
 kinematical parameters  for  $\gamma^*(q)+N(p)\rightarrow \gamma^*(q)+N(p)$
we note that

\be
s-m_N^2=Q^2\left(\frac 1x-1\right)
\ee
with  Bjorken-x defined as $x=Q^2/2p\cdot q$ and  $Q^2=-q^2\geq 0$.
In the Pomeron regime with  $s\gg Q^2\gg -t$,  we have $s\approx Q^2/x$. From the identification (\ref{RAPID})
it follows that for fixed $Q^2$

\be
\label{CHIX}
\chi={\rm ln}s={\rm ln}\,n={\rm ln}\,\frac 1x
\ee
The larger the boost, the larger the rapidity, the smaller the range of Bjorken-x
probed by the string transverse fluctutations. 
Hence,  $x=\frac 1n$ is  identified as  the fraction of longitudinal momentum 
carried by each of the  transverse n-string bits.

 It follows that the  string fluctuations as  wee-quanta carry longitudinal momentum, where
the  mean captured by the  $F_2(x)$  structure function at low-x is ($n\gg N$)

\be
F_2(x)\sim\sum_{l=0}^\infty  lp_l[D_\perp,N,n,1]=D_\perp\,n^\Delta=\frac{D_\perp}{x^\Delta}
\ee
The exponent $\Delta$ is fixed by  the zero point motion or quantum entanglement 
of the string through the Pomeron intercept

\be
\Delta=\frac 16\rightarrow \frac 2{D_\perp}
(\alpha_{\mathbb P}(0)-1)
\ee
Recall from (\ref{SATXX}) and the ensuing estimate, that the entanglement entropy saturates for $\chi_S=10.6$.
Using (\ref{CHIX}),  this translates to a saturation at low-x when  $x_S\approx 2\,10^{-5}$.

\section{Conclusions}

In walled AdS, parton-parton scattering at large $\sqrt{s}$ is dominated by 
the exchange of a hyperbolic surface that Reggeizes through 
a worldsheet instanton. The zero point motion of
the string is characterized by a quantum or thermodynamical
entropy ${\cal S}_T$ that is tied to the rise of the scattering amplitude
and multiplicities in hadron-hadron scattering at large rapidities.

The surface is spatially entangled with an entanglement
or von-Neumann entropy ${\cal S}_E$ that coincides with the quantum or
thermodynamical entropy  ${\cal S}_T$, i.e. ${\cal S}_T={\cal S}_E$. This 
entanglement
entropy coincides with 
the one in critical 2-dimensional conformal field theories and spin chains with a central charge 
$D_\perp$. 

At asymptotic rapidities, the ratio of the entanglement entropy to the
transverse area of the string is similar to that of a black hole. The string appears
maximally entangled with a saturation momentum fixed by the string length.
This suggests that the  rate of growth of the entanglement entropy when cast in 
terms of Gribov diffusion time, is at the chaos bound. These observations
maybe at the origin of the fast scrambling of information  
and  collectivization  in pp collisions  as recently 
 reported by the CMS collaboration~\cite{CMS}, and argued in~\cite{ED}. 

The largest eigenvalues of the 
entangled density matrix obey a cascade equation in rapidity. They describe 
the  probability distributions of wee-quanta at low-x and strong coupling, much
like the BFKL wee-dipoles at weak coupling. They are
measurable through the multiplicities in hadron-hadron scattering or structure
functions in  deep-inelastic scattering  at present or future colliders.

\section{Acknowledgements}
This work was supported by the U.S. Department of Energy under Contract No.
DE-FG-88ER40388.

\newpage


 \vfil

\end{document}